Review

Leonardo Andres Espinosa Leal and Olga Lopez-Acevedo*

# On the interaction between gold and silver metal atoms and DNA/RNA nucleobases – a comprehensive computational study of ground state properties

**Abstract:** The interaction between metal atoms and nucleobases has been a topic of high interest due to the wide scientific and technological implications. Combining density functional theory simulations with a literature overview, we achieved an exhaustive study of the ground state electronic properties of DNA/RNA nucleobases interacting with gold and silver atoms at three charge states: neutral, cationic, and anionic. We describe the nature of the stability and electronic properties in each hybrid metallic structure. In addition to the metal interacting with the five isolated nucleobases, we included their respective DNA-WC base pairs and one case with the protonated sugar-phosphate backbone. As a general trend, we discerned that the energetic ordering of isomers follows simple electrostatic rules as expected from previous studies. Also, we found that although the metal localizes almost all of the extra charge in the anionic system, a donation of charge is shared almost equally in the cationic system. Furthermore, the frontier orbitals of the cationic system tend to have more effects from the pairing and inclusion of the backbone than the anionic system. Finally, the electronic gap varies greatly among all of the considered structures and could be further used as a fingerprint when searching DNA-metal hybrid structures.

**Keywords:** DNA-metals; gold and silver; nucleobases.

***Corresponding author: Olga Lopez-Acevedo,** COMP Centre of Excellence, Department of Applied Physics, Aalto University, P.O. Box 11100, 00076 Aalto, Finland,
e-mail: olga.lopez.acevedo@aalto.fi
**Leonardo Andres Espinosa Leal:** COMP Centre of Excellence, Department of Applied Physics, Aalto University, P.O. Box 11100, 00076 Aalto, Finland

## 1 Introduction

The main workhorses for the characterization of the electronic chemical and physical properties of matter in terms of *first principles* methods are the Post-Hartree-Fock methods and density functional theory. After a tremendous development of the molecular orbital theory [1–3], the first successfully applications for the description of the electronic structure of complex molecular biological systems appeared early in the seventies with the works of J.A. Pople [4–7] and E. Clementi [8–10]. However, the inaccuracy on the developed exchange-correlation functionals tilted the balance toward the wavefunction-based methodologies, in particular, the Møller-Plesset perturbation theory [11]. In general, these methods scale poorly with the size of the studied systems; therefore, small basis sets had to be used, or some constraints were necessary to obtain feasible results within the computational resources available at that time [12–14]. Owing to these restrictions, the studies were mostly limited to organic systems interacting with mono- or bivalent metals with a low number of electrons [12–17].

The study of the interaction of biological systems with transition heavy metals, and in particular the nucleic acids, had to wait until the development of more complete specific basis sets [18–20] that are able to predict with accuracy the well known experimental data. Almost simultaneously, the apparition of better and more sophisticated exchange-correlation functionals, such as local, gradient-corrected, or the highly successful hybrids, changed the map of methodologies in terms of computational calculations [21–23]. These functionals allowed to use larger basis because of the better scalability of the method; therefore, less computational resources were required in order to have better values for comparing with experiments.

The aim of this review is to provide a framework useful for the study of the relationship between DNA and transition metals and, in particular, to fill all the missing characterization of the interaction of nucleobases with gold and silver using the first principles methods. In the manuscript, we have first included a historical context that follows the path of the different scientific outcomes, and then, we have recapped the main conclusions obtained by comparing the available scientific literature and our own calculations using a DFT/real-space methodology (more details in the Section 2).

## 1.1 Metal-DNA interaction: historical path

DNA is a complex polymer with several structural levels classified from primary to quaternary. In the primary level, the ordering of nucleobases aligned and attached are described. The secondary level includes the ordering between two bonding nucleobases, the tertiary level describes their three-dimensional spatial organization, and there is a last quaternary level high-level structure related to its organization with respect to other molecules [24]. If there were a molecule equipped with better properties than DNA, probably evolution would have selected it for the important task of storing and encoding the genetic information of the living beings on Earth. Despite all the work and effort invested by the society and the scientific community, there are still remaining questions concerning the nature, stabilization, self-recognition, self-assembly, hybridization, replication, and interaction, among others.

Interaction of metal ions with DNA has been a topic of extreme importance [29] since the discovery of the structure of this molecule by Watson and Crick in 1953. Owing to the charged nature of the phosphodiester backbone, the stabilization of the long strands in solution strongly depends on the presence of metallic cations [26, 30]. It has been shown experimentally that at high salt concentration, the DNA duplex is more stable and that this stability depends on the composition of the added salt [31, 32]. Moreover, one of the revolutions in the study of metal-DNA structures came with the discovery of *cisplatin* as chemotherapy drug for cancer treatment [33]. The cisplatin binds the DNA through the nucleobases inhibiting the replication mechanisms and additionally producing *apoptosis* in the affected cells (see review by Sherman and Lippard [34] and references therein). This breakthrough gave a huge impulse in the search of alternative compounds with transition metals as prospects for anticancer treatment in general, in particular, because the cisplatin can also lead to side effects on other types of healthy cells [35]. Therefore, it is of great importance to design more effective and less toxic drugs [36, 37]. Another topic of great interest is the possibility to induce and control magnetic or electronic properties on metal-DNA/RNA complexes (see minireview by Clever et al. [38] and references therein) for storage purposes. Finally, but not less important, in the last years, a new phenomena has appeared to expand the scientific horizons: hybrid DNA-metallic nanostructures present well-defined and tunable optical properties depending on the size and nature of both interacting structures. In isolated gold and silver metallic structures, the plasmonic effects are a well-known phenomena [39, 40] (see book edited by Chris Geddes [41] and references therein); therefore, it is possible to extend it to more complex arrangements to build DNA-plasmonic nanostructures [42, 43]. Moreover, at low metal concentrations, it has been found that there is the appearance of high fluorescent properties upon interaction with few-atom noble metal clusters, in particular, gold [44] and silver [27, 45]. Recently, an experimental breakthrough has made it possible to measure the composition of DNA-stabilized fluorescent silver clusters [28]. These advances in separation techniques and optical characterization have led to the first identification of numbers of neutral silver atoms, silver ions, and DNA strands contained in fluorescent Ag-DNA complexes. Based on their results, the experimental group has proposed several models of highly charged rod-like structures that would follow a shell (or jellium model) for absorption and emission properties [46] (See Figure 1 for some representative examples of metal-DNA structures).

Under this context, the study of single isolated atoms interacting with DNA/RNA is the first step for a bottom-up understanding of the underlying mechanisms and properties during the formation of stable metal isolated or mediated DNA/RNA strands. In particular, the importance for the study of the geometry of monocationic silver cytosine-cytosine [47] and dicationic mercury thymine-thymine [48] strands has been shown.

The first computational works on the interaction of cationic metals with DNA-nucleobases were published in 1983 by K.P. Sagarik and B.M. Rode from the University of Innsbruck [12–14]. Using the freezing geometries of single nucleobases and DNA Watson-Crick (WC) base pairs obtained from the crystallographic experimental data, the authors studied the interaction of monocationic lithium and sodium as well as dicationic magnesium and calcium atoms. Their results show the reactive sites for the respective ions and the stability of the hydrogen bonds in the base pairs under the cationic interaction. One year after the seminal works of Rode, Del Bene [15, 16] (former posdoct of Pople), Hobza and Sandorfy [17] discuss the case of the single metallic cation on DNA base pairs without any symmetry constraints. In the first, a paper from the series of works on the *Molecular orbital theory of the hydrogen bond* (a beautiful and complete series of 28 communications about this topic that begins in 1971 [49] and ends in 1985 [50]), the author studied the interaction effect of the lithium monocation on WC base pairs: AT and GC. In the second, the authors study only the AT pair but including the effect of different mono- and dications on the structure. In both cases, the computational methodology used was HF/STO-3G.

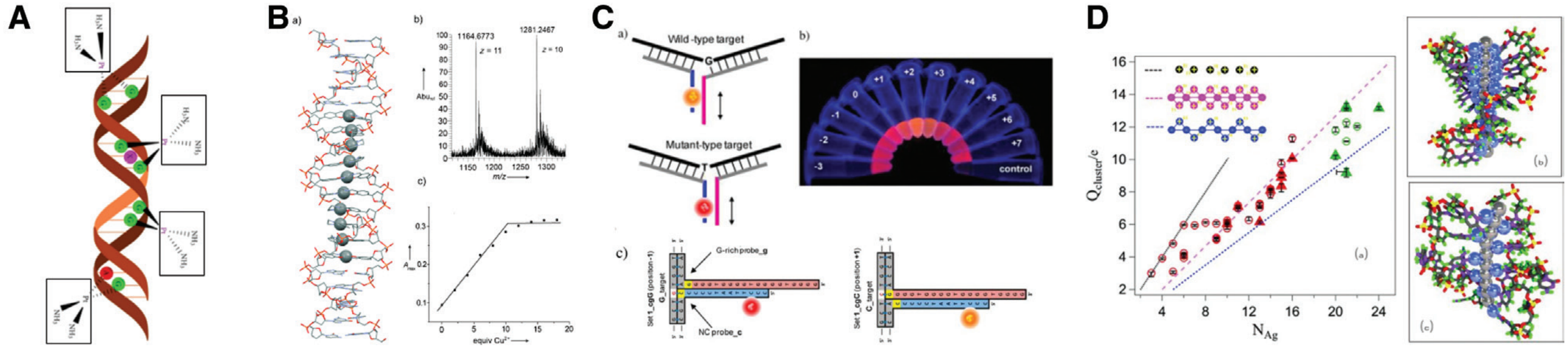

**Figure 1:** Different metal-DNA molecular structures with important applications: (A) DNA adduct formation with cisplatin showing two amino groups coordinated on the platinum atom. From Ref. [25]. (B) A helical turn full of metal ions. The metal-salen base-pair concept allows the incorporation of up to 10 transition-metal ions into double strands of DNA (see picture; Mn gray spheres, C gray, N blue, P orange, O red). The crosslinking provided by the metal-salen complexes conveys a high stability to the self-assembled systems. From Ref. [26]. Reprinted with permission of John Wiley & Sons, Inc. (C) Chameleon NanoCluster Beacon lights up into different colors upon binding SNP targets. The emission of color changes depending on the nature of the target. The light-up color of NanoCluster Beacons can be tuned by repositioning the enhancer sequence with respect to the NC-nucleation sequence. Adapted with permission from Ref. [27]. Copyright (2012) American Chemical Society. (D) Silver cluster charge, Qcl, plotted versus the number NAg of silver atoms (neutral and ions) in fluorescent (solid triangles) and dark (open circles) Ag:DNAs and examples of the corresponding structures. Adapted from Ref. [28] with permission of John Wiley & Sons, Inc.

Calculations with more accurate basis sets in combination with relativistic effective pseudopotenials, in particular, for transition metal atoms, were published in 1996 and 1997 by Burda, Šponer, Leszczynski, and Hobza. Their investigations use a HF/6-31G**/DZ methodology for the geometrical optimizations and the second-order Møller-Plesset with the same basis set for the interaction energies. For the first time, single nucleobases [51] and WC [52] base pairs were studied interacting with gold and silver monocations. During the next years, the same authors continued working on the same topics, and these pioneer works were summarized in the theoretical review on the subject titled, *Electronic properties, hydrogen bonding, stacking, and cation binding of DNA and RNA bases* [53]. Interestingly, in the introduction, the authors pointed out the incursion of the DFT methods in the field. Nonetheless, contrary to their predictions, the computational methods based on the electronic density would dominate during the first decade of the 21st century, and only in the recent years the wavefunctions methodologies are making a come back to compete directly with DFT in accuracy and performance [54, 55].

The modern quantum chemistry calculations on metal-DNA/RNA systems have been ruled by following the DFT/B3LYP/LANL2DZ recipe. The first works on the topic were carried on by Kryacho et al. [56–62]. In their first paper, the authors regard on the properties of small neutral gold clusters with isolated nucleobases. In the case of silver, the seminal work was done by Vrkic et al. [63]. Here, the authors studied the binding sites for cationic silver in adenine. In the following years, several computational studies appeared; these mainly focused on DNA/RNA-gold/silver hybrid structure properties by fixing the charge state (neutral, cationic, or anionic) [64–66] or by considering a reduced number of nucleobases [67, 68] (or base pairs) [69] in gas phase. In recent years, with the need to understand the aforementioned phenomena in more complex systems, the investigations in hybrid DNA/RNA metallic nanoclusters (MNCs) have opened new fields of research such as the interaction with large clusters [70, 71], expanded nucleobases [72–74], influence of solvent [75], importance of the sugar backbone [76, 77], among others. Furthermore, the rapid increase in the size of the systems or the nature of the phenomena (*vide supra*) have attracted the attention of other alternative methodologies with better performance such as real space [64], molecular dynamics [78] or mixed QM/MM [77].

After almost two decades on computational research of the electronic properties of DNA/RNA interacting with gold and silver, we present in this manuscript the first exhaustive review on the *ab initio* computational area. We have collected from the available scientific literature the main results, and we have compared them with our own calculations. In addition, we completed the missing data. The main goal is to summarize and discuss in one single document all the relevant information toward a full description of the ground state properties of the hybrid DNA/RNA-Au,Ag systems.

This review is organized as follows: In Section 2, we describe the computational methodology used in our calculations and we present the analyzed structures, in Section 3, we summarize and discuss our results comparing them with other computational studies available in the literature, and in Section 4, we give the conclusions and perspectives on this field of research.

## 2 Computational methodology

In this work, we provide an overview of every possible gold and silver metal atom in DNA/RNA geometries – reporting each atoms corresponding to electronic properties as a step toward the modeling of hybrid DNA/RNA-MNCs. We carried out this research using the simplest model of a single nucleobase – guanine (**G**), adenine (**A**), thymine (**T**), cytosine (**C**), and uracil (**U**) – interacting with one noble metal atoms of gold (Au) and silver (Ag) at different charge states: cationic (+1), neutral (0), and anionic (-1). We studied structural and electronic properties, including ionization potential and electron affinity, Bader charge, electronic gap, and localization of frontier orbitals. For every property, the effects of pairing between the WC base pairs: guanine-cytosine (**GC**) and adenine-thymine (**AT**), and the presence of the protonated sugar-phosphate backbone **dGMP** (guanosine nucleotide) were also simulated and discussed.

There are many others that are important and, in consequence, interesting geometries associated to the topic of the interaction of DNA/RNA with gold and silver such as tautomers, non-WC base pairs (Hoogsteen, Wooble, homopolymers) or negative-charged sugar-phosphate backbone. However, their study are out of the scope of this review, and we encourage the reader to fill this gap regarding the available scientific literature. For this review, we have chosen to deal only with the canonical forms of nucleobases and WC base pairs.

We studied the following systems: DNA/RNA nucleobases (see *top* of Figure 2), the DNA WC base pairs (see *center* of Figure 2), and one nucleotide with guanine as nucleobase: dGMP (see *bottom* of Figure 2). The work occurred in two parts. In the initial stage, we performed the geometrical global optimizations using the Basin Hopping algorithm [79, 80] without any symmetry constraint. A different set of initial configurations were used for each system to find the minimal geometries. In a second stage, we characterized the obtained structures in the ground state: binding energies, electronic gaps, and charge analysis using the Bader method [81].

All calculations in this work have been performed using the DFT code *GPAW* [82, 83], which combines real space methods and the projector augmented-wave formalism [84]. We used 8.0 Å of vacuum around the systems in a box-shape simulation box and a real-space grid spacing of 0.18 Å in all simulations. The structure relaxations were performed until the atomic forces were below 0.02 eV/Å. The electronic states H(1*s*), C(2*s*2*p*), N(2*s*2*p*), O(2*s*2*p*), Ag(4*p*4*d*5*s*), Au(5*d*1*s*) were included as valence states, whereas the core states were frozen. The setups were built up, taking into account the relativistic correction for the metal noble atoms.

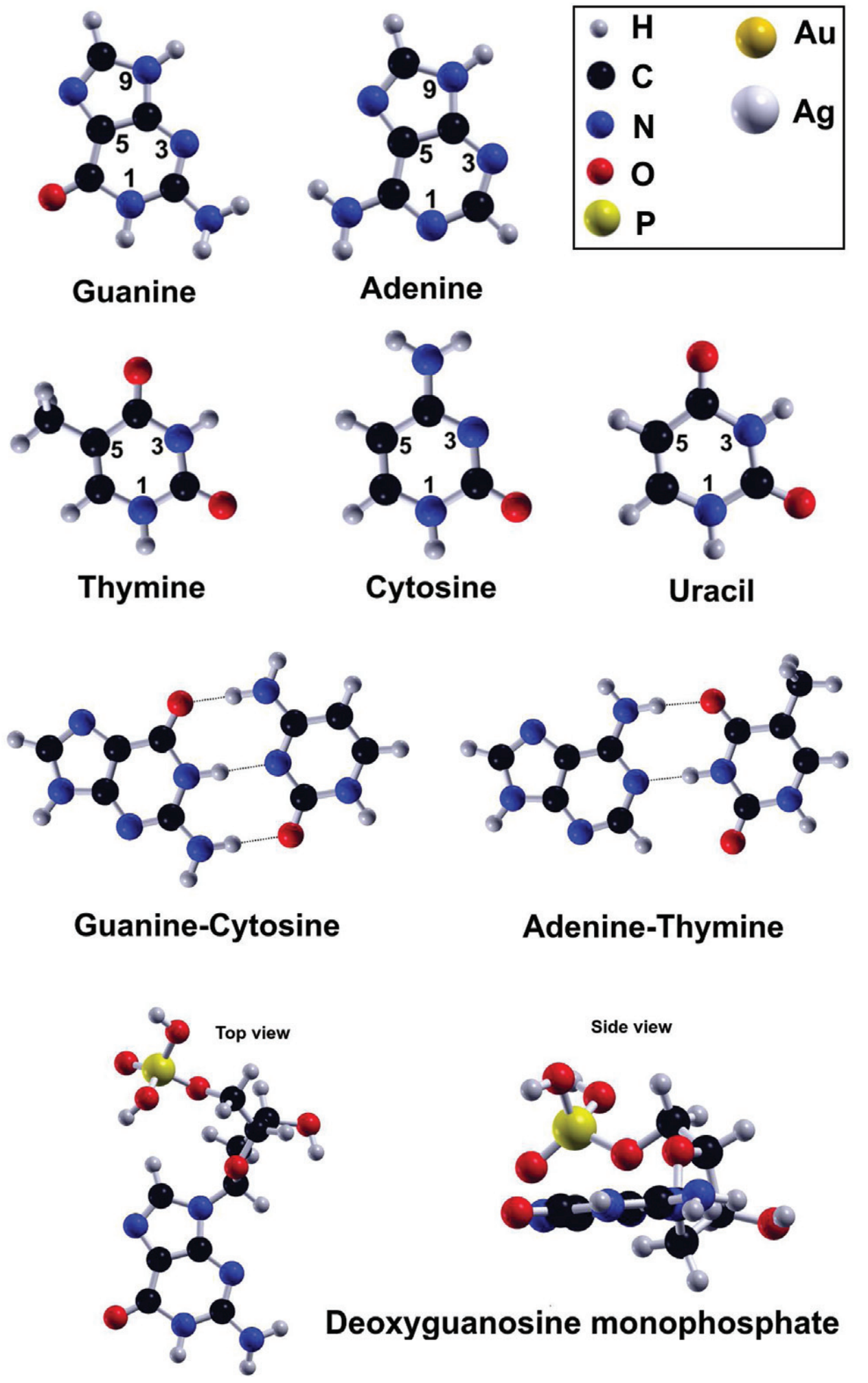

**Figure 2:** Structures studied in this work. *Top:* Five protonated nucleobases. Guanine (**G**) and adenine (**A**) (purines) and thymine (**T**), cytosine (**C**) and uracil (**U**) (pyrimidines). *Center*: Protonated DNA WC base pairs: guanine-cytosine (**GC**) (*left*) and adenine-thymine (**AT**) (*right*). *Bottom*: Top view (*left*) and side view (*right*) of the guanine nucleotide: deoxyguanosine monophosphate (**dGMP**). In the top-right box the atomic color convention.

In the configurational search, a temperature of 1000 K and a maximal step-width of 0.1 Å were combined with the limited-memory Broyden-Fletcher-Goldfarb-Shanno (LBFGS) algorithm [85] as the local optimizer for the Basin Hopping algorithm. All the RNA/DNA geometries were built using the 3DNA software package [86] and manipulated using Avogadro [87]. The figures of structures and orbitals were plotted using XCrySDen [88]. All software used in this work is freely available for download in the Internet. GPAW, Avogadro and XCrySDen are released under the GNU General Public License. 3DNA is free of charge for non-commercial use without any warranty.

## 3 Results and discussion

Several studies have shown that the DFT methodology is accurate enough to describe the geometry and the energy bindings of the hydrogen bond interactions in DNA structures [90–92]. Therefore, the first step in our research was to choose the first principles exchange-correlation functional used in all the calculations. We compared the performance of three exchange-correlation functionals: A LDA (PW) [93], GGA (PBE) [94], and an improved version of the revised PBE (RPBE) [95]. The systems used as test cases were, without loss of generality, both guanine-Au and guanine-Ag structures with the same initial configurations at the three charge states (for more details, see Supplementary Information). For all hybrid metallic DNA structures, the binding energy was determined by subtracting the total energy of the composed system to the sum of the individual energy of each moiety. For all structures, it is assumed that the charge (added or removed) remains in the metallic atom.

We found that all three exchange-correlation functionals predict in the same order of energy, the first stable structures. However, LDA tends to give the largest number of bonding sites. The binding energies and bond lengths for the lowest energetic Au/Ag-guanine structures are consigned in the Table 1. The local functional predicts a higher binding energy in comparison to PBE and RPBE (binding energies: PW>PBE>RPBE). On the other hand, the RPBE produces lower values in the binding energies as opposed to other studied DFT methodologies, but in general, it gives the largest bond lengths. Meanwhile, the PBE method gives values for the binding energies and for the bond lengths in between LDA and RPBE (bond lengths: RPBE>PBE>PW). After carefully reviewing our results, it was apparent that the binding energies were shifted from PBE to RPBE for a constant value; therefore, we decided to use PBE for all of our calculations (see Supplementary Information). A comparison with other gradient-corrected theoretical calculations[66, 96] show that the election of PBE as our DFT exchange-correlation approach is the most advantageous choice because it can be extensive to study larger DNA-metal structures with good accuracy, in particular, for charged systems, without a huge increase in the computational load. In the subsequent subsections, the computed values will be compared to values from accurate quantum chemistry methods supporting this choice.

### 3.1 Structures

The interaction of gold and silver atoms with DNA/RNA nucleobases presents a heterogeneous behavior; however, using our calculations, we extracted an important set of features about the interactions. In general, we can describe the dynamic of the bond between one metallic system and the nucleobases through the analysis of the charge distribution in the molecular structure. In the isolated molecule, the hydrogen present in the imide (N-H), amine (N-$H_2$), and alkene (C-H) functional groups tends to donate a charge to its bonded atoms, nitrogen and carbon, by following the order of electronegativity: N (3.04)>C (2.55)>H (2.20). Additionally, due to the high electronegativity of oxygen (3.44), this also can act as an

**Table 1:** Binding energies (bond length) in kcal/mol (in Å) for the most stable hybrid guanine-Au/Ag structures in the three charge states: neutral, cationic and anionic (*n*: 0, +1, -1) obtained with three different exchange-correlation functionals: PW (LDA), PBE and RPBE (GGA).

| | Charge (*n*) | PW (LDA) | PBE (GGA) | RPBE (GGA) |
|---|---|---|---|---|
| Au | 0 | 27.5 (2.15) | 13.38 (2.26) | 7.19 (2.33) |
| | +1 | 116.07 (1.99) | 95.92 (2.04) | 87.48 (2.06) |
| | -1 | 40.96 (2.13) | 31.28 (2.20) | 27.77 (2.29) |
| Ag | 0 | 15.15 (2.24) | 6.30 (2.45) | 2.79 (2.64) |
| | +1 | 92.59 (2.18-2.32) | 75.07 (2.26-2.43) | 67.93 (2.31-2.51) |
| | -1 | 35.31 (2.20) | 28.02 (2.37-2.77) | 25.02 (2.51) |

The binding energies are calculated for a given value of charge as $E_b$=E(base)+E(metal)$^n$-E(base+metal)$^n$.

important bonding site for the metallic atoms. The redistribution in the electronic potential creates different zones in the nucleobase where the hydrogens behave like positive centers of charge and the rest as negative centers of charge (see Figure 3 for reference).

This fact defines the nature of the interaction of the nucleobase with external charged systems. In general, the binding energies for the systems that interact with silver are much lower than gold; they correlate with a larger bond distance. For most of the obtained hybrid metal structures, the symmetry was planar, with a few exceptions in the isolated cationic nucleobases – in particular, in the interaction with the silver atoms. More exceptions appear in the cationic base pairs, and when the sugar backbone is considered, the planarity becomes the exception.

Next, we discuss the main structural features obtained from our calculations using a DFT/PBE methodology, ordering the interactions from most to least stable, and listing the preferred binding sites in order of stability (see the most stable hybrid DNA/RNA structures for both gold and silver in the Figures 4 and 5, respectively. For the complete set of structures, see the Supplementary Information).

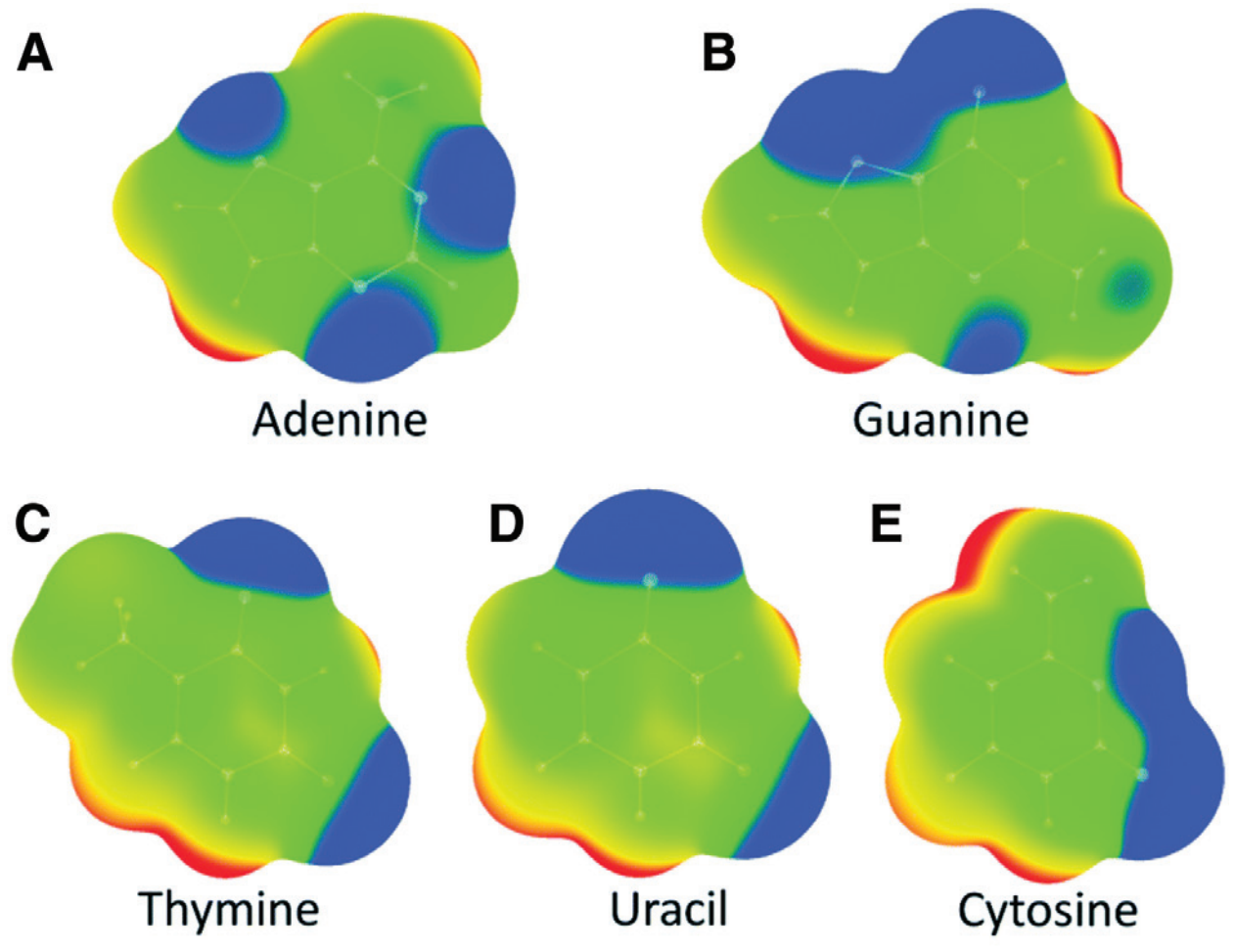

**Figure 3:** Computed MP2/cc-pVTZ electrostatic potential on the 0.001 au surface of (A) adenine, (B) guanine, (C) thymine, (D) uracil, and (E) cytosine. Color ranges, in kcal mol$^{1}$, are given as follows: red, >31.31; yellow, from 10.17 to 31.31; green, from -10.92 to 10.17; blue, more negative than -10.92. In all nucleic bases, hydrogen atoms from C–H fragments have positive potentials. From Ref [89]. Reprinted with permission of the Royal Society of Chemistry.

### 3.1.1 Neutral nucleobases

In principle, the neutral metal atom can interact with both the positive and negative centers. The variety of structures and type of bonds in these systems can be explained due to the hybridization of the *s*-orbital with the *d*-orbital in the noble metal atom (see the orbital analysis in Section 3.2.4). It allows a redistribution of the electronic cloud that creates an effective dipole moment that interacts with both the positive and negative centers in the molecule. This fact explains in a simple manner the previously reported *nonconventional hydrogen bonds* [56, 57] present in the interaction of bare neutral clusters with DNA/RNA nucleobases. The obtained values for the binding energies show that gold atoms are energetically more stable than silver when these interact with the nucleobases. The binding energies for the Au-nucleobase ranges from 13.38 to 7.70 kcal/mol (in the most stable structures) obeying the following order **G**>**C**>**A**>**T**>**U**. In the case of Ag-nucleobase, the binding energies ranges from 8.12 to 3.80 kcal/mol (in the most stable structures), and they obey the order **C**>**A**>**G**>**T**>**U**. Using B3LYP/6-31(d,p)/ LANL2DZ calculation method, Martínez et al. [67, 68] have found the same geometries for the neutral Au/Ag-uracil structures; however, the order of stability differs with our calculations. This results can be explained because all the hybrid structures are very close in energy. The most stable geometry obtained for the Au-adenine in our calculations is the same. The dissociation energies obtained for this structures are in the same order of magnitude and follows the order to the results in the references: **Au-A** (10.3 kcal/ mol)>**Au-U** (4.9 kcal/mol) and **Ag-C** (7.9 kcal/mol)>**Ag-U** (2.9 kcal/mol). It is important, pointed out here, that in the paper by Kryachko and Remacle [56], the binding energies that the authors describe in the introduction for the gold neutral single-nucleobases are lower than 5 kcal/mol, but with any detail about the calculations. In the work by Acioli and Srinivas [66], the authors study the interaction of neutral silver and gold atoms with nucleobases. Their approach use the BPW91 exchange-correlation [97] functional in combination with the Stuttgart effective core potentials to describe the valence electrons of gold and silver atoms and the all-electron DGTZVP basis set for rest. Their results show a binding energy order for the gold as follows: **A**>**C**>**G**>**T** and for silver **C**>**A**>**G**>**T**. The respective range of values are 11.2–3.8 kcal/mol for gold and 5.3–3.3 kcal/mol for silver, in good agreement with our calculations.

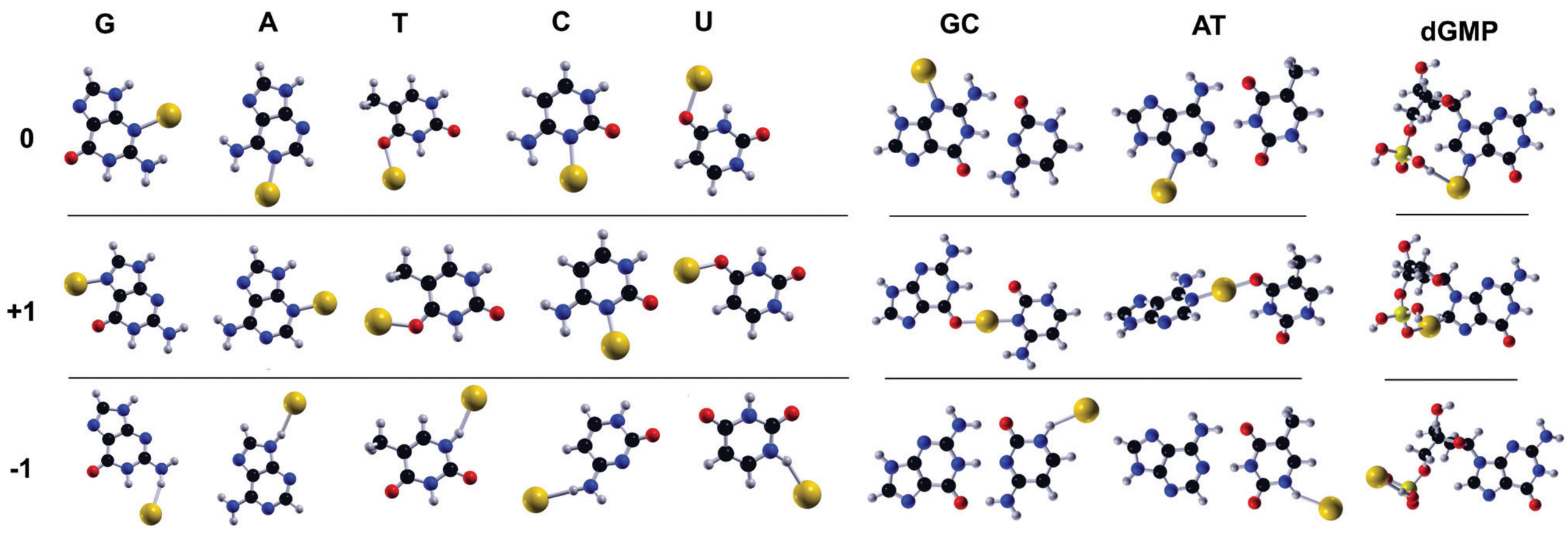

**Figure 4:** Most stable hybrid gold-DNA/RNA structures (nucleobases, WC base pairs, and deoxyguanosine monophosphate) for the three charge states.

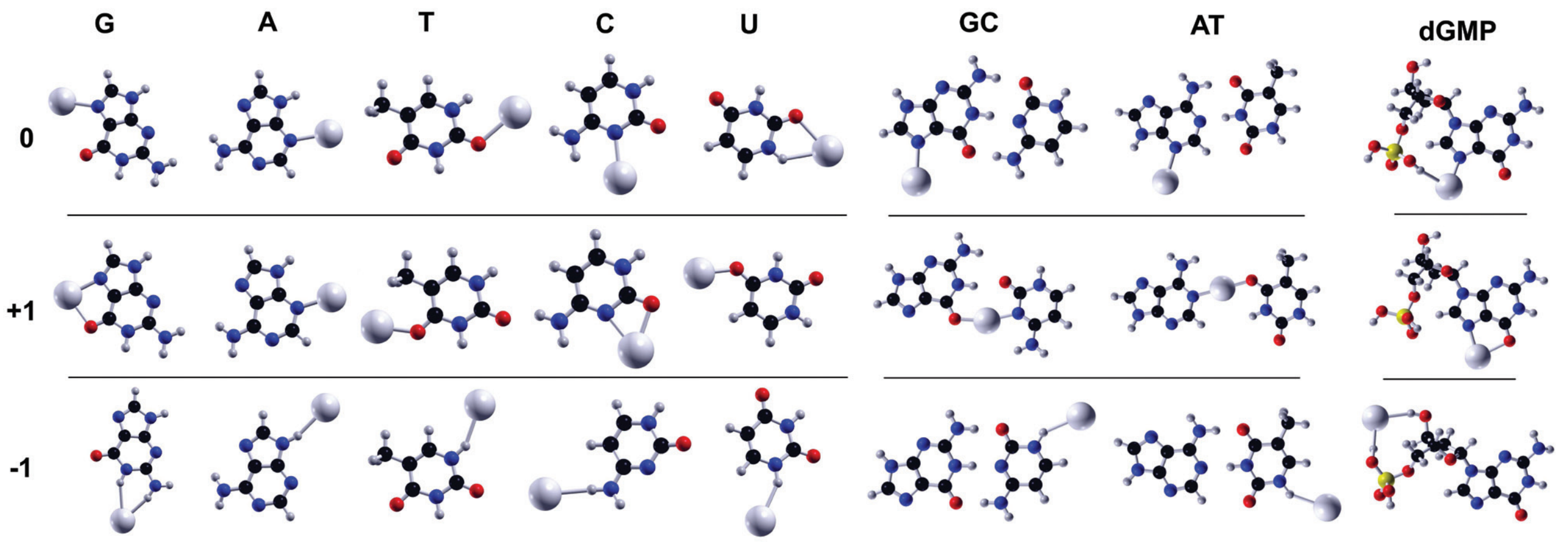

**Figure 5:** Most stable hybrid silver-DNA/RNA structures (nucleobases, WC base pairs and guanine monophosphate) for the three charge states.

On average, the binding energies for gold are around twice the values obtained for silver. The bond lengths for the gold bases are around 0.2 Å shorter than silvers. This conclusion is supported by other theoretical calculations with different methodologies on the same kind of structures [66–68]. The first preferred binding sites for the guanine follows the order N3>N7>O6 for gold and N7>O6>N3 for silver. Additionally, in the case of the bond with the oxygen, there is some interaction with the hydrogen in NH1. The Au-adenine follows the order N1>N7>N3, and the Ag-adenine N3>N1>N7. The metal-thymine hybrid structures follow the order O4>O2-NH1>O2-NH3 for the gold and O2-NH1>O4>O2-NH3 for the silver. In cytosine, the preference for the binding sites follows the order N3>O2-NH1>O2 for both the gold and silver. The metal-uracil systems obey the order O4-NH3>O2-NH1>O4-CH5>O2-NH3 for the gold and O2-NH1>O4-NH3>O4-CH5>O2-NH3 for the silver.

### 3.1.2 Cationic nucleobases

In the nucleobase, the hydrogen atoms are charged positively; therefore, they interact repulsively with the cationic metal. The strongest attractive interactions are done through the N and O atoms or in a combination of both when they are adjacent in the molecular structure (guanine and cytosine). This fact can be confirmed by means of the Bader analysis (see Subsection 3.2.2). The binding energy in cationic structures is higher in comparison to the neutral and anionic systems. In the lowest energetic cationic structures, the energy bindings range from 95.92 to 67.97 kcal/mol in the hybrid cationic gold-nucleobases and from 75.07 to 48.16 kcal/mol in the silver systems. The first case obeys the order **G**>**C**>**A**>**U**>**T** and the second one follows **G**>**C**>**A**>**T**>**U**. The bond lengths in the gold species are, on average, around 0.15 Å shorter than the silver-nucleobase structures. The special cases of the binding of cationic structures are the guanine and the cytosine, these have the highest binding energies, this effect is caused due to the availability of both adjacent nitrogen and oxygen as binding sites for the cationic metal. This fact is of high importance in the stabilization of cationic-mediated DNA homopolymers or for design antitumor drugs able to bind in these specific sites.

A comparison with other theoretical results [67, 68] for cationic gold show a good agreement in energies and stabilization order. The authors found a relative order (binding energy) **A** (80.8 kcal/mol)>**U** (62.0 kcal/mol), the discrepancy with our results is on average 7.1 kcal/mol. From the same authors [98], the interaction with the cationic silver atom shows the order **C** (65.9 kcal/mol)>**U** (43.9 kcal/mol) and the discrepancy respect our calculations is, on average, only 3.4 kcal/mol. Most accurate calculations [99] using a MP2/cc-pVTZ/MWB level of theory shows a binding energy for the cationic Ag-adenine of 53.1 kcal/mol; this value compares very well with our result of 57.99 kcal/mol. For the Ag-cytosine cation, the same authors reports a value of 61.9 kcal/mol, we obtained a binding energy of 68.50 kcal/mol.

For the cationic systems, we can summarize our results as follows: the preferred binding sites for the cationic gold-guanine systems follow the order N7>N3>O6 and for the silver N7-O6>O6>N3>N3>N1, with these last two lying out of the plane of the molecule. For the gold-adenine, the binding sites obey the order N3>N1>N7, which coincides with the binding sites for the silver-adenine systems. In the case of the hybrid gold-thymine structures, the binding sites follow the order O4>O2 (H3)>O2 (H1) (the sites in parenthesis indicate the orientation, but not a physical-chemical binding), and for the silver-thymine structures, the order O4 (C5)>O2 (H3)>O4 (H3)>O2 (H1). The interaction of the cationic cytosine with the gold metal in the cationic state obeys the order N3>O2 (N3)>O2 (H1), and in the case of the silver metal, the order is N3-O2>O2 (H1). For the case of the cationic gold-uracil structures, the binding sites follow the order O4 (H5)>O4 (H3)>O2 (H3)>O2 (H1), and for the hybrid silver-uracil, the order is O4 (H5)>O4 (H3)>O2 (H3) nonplanar>O2 (H3)>O2 (H1). The binding sites and the order are almost the same, except in the case of silver, where one intermediate non-planar structure appears.

### 3.1.3 Anionic nucleobases

The interaction of the nucleobases with the gold atom in the anionic system presents a bond length (in average) 0.3 Å shorter than in silver. The binding energy of the most stable structures in gold is around 4 kcal/mol larger than in the equivalent silver system. The topology of the hybrid metal-nucleobase structures is influenced by the exclusive electrostatic attraction with the hydrogens in the base and the strong repulsion by the O, N, and C atoms. This is because the noble metal is almost totally charged negatively (see the Bader analysis in Section 3.2.2). The order of stability in the metal-base structures is the same for both gold and silver: **G**>**T**>**U**>**C**>**A**. The binding energies range between 26.03 and 31.28 kcal/mol for gold and 22.24–28.02 kcal/mol for silver. For the gold-guanine structures, the binding sites follow the order H2 (H1)>H9>H2 (N3), and for the silver case, the order of binding is H2-H1>H9, with a double bond for the most stable in the last case. For the gold-adenine structures, the order of stability is H9>H6 (N1)>H6 (N7), and for the silver-adenine, we have the same first two sites: H9>H6 (N1). In the thymine-obtained structures, in the case of the interaction with gold, the order of stability in the binding sites follows H3>H1, and for the interaction with silver, the only place is H3. For the cytosine structures, the binding sites follow the same order for both gold and silver H4 (H5)>H1>H6 (this last one is a particular structure when the hydrogen is bonded to a carbon atom). In the gold-uracil structures, the order in the binding is H1>H3, and in the case of the of the silver atom, the only binding site is H1.

A direct comparison of our results with the available literature shows a good agreement in the binding energies for the anionic species. Valdespino and Martínez [68] predict for the gold hybrids **A** (26.1 kcal/mol)>**U** (24.8 kcal/mol) and for the silver case **U** (22.4 kcal/mol)>**C** (20.7 kcal/mol). In the work of Cao et al. [96], for the single gold anionic structures, the order of stability predicted is **U** (22.6 kcal/mol)>**G** (21.68 kcal/mol)>**T** (20.99 kcal/mol)>**C** (19.14 kcal/mol)>**A** (16.14 kcal/mol). The order in the stability is different for the first three nucleobases; however, the range of energy is the same of our calculation, but shifted around 10 kcal/mol.

### 3.1.4 Watson-Crick base pairs

The effect of pairing between two DNA nucleobases give origin to a big set of stable structures when these interact with a noble metal. This number of structures strongly depends on the charge in the metallic atom, and the structures can form new geometrical distributions. For reference, our results for the binding energies at PBE level for both free-metal WC **GC** and **AT** base pairs are 25.45 kcal/mol and 13.49 kcal/mol, respectively. The obtained values are in relative good agreement with other theoretical results obtained using more accurate methodologies such as MP2 [100, 101] or dispersion-corrected functionals [102, 103].

With respect to these methods, the discrepancy in the binding energy runs between 1% and 12%. These values in the binding energy can change due to the interaction with the metallic atom (charged or neutral). This effect will be described in the next paragraphs.

In both hybrid metal base pair structures, the stability for the lowest structure in three charge states follows the rule Au/Ag-**GC**>Au/Ag-**AT** (with Au-pair>Ag-pair). For the neutral and anionic cases, in the lowest energetic structure, the most stable structures are those where the metal atom binds far from the binding region between the two nucleobases (WC region). For the first set, this behavior is related with the tendency of the hybrid structures to keep the geometry of the base pair. In the anionic, this is due to the strong electrostatic repulsion of the hydrogen centers with the negatively charged metal. For the cationic, the most stable structures are those ones where the metal binds between the two nucleobases in the region of the hydrogen bridges. It is important to highlight here that in our study, we kept the WC conformation, even if we do not restrict the search of structures, during the search of minimal structures. Therefore, the founded base pairs geometries mostly follows this conformation.

The hybrid neutral gold-**GC** base pair structures obeys the order of stability **G**(N3)>**G**(N7)>**C**(O2-NH1)>**G**(O6)>**G**(O6)-**G**(NH1)-**C**(N3)>**G**(NH2)-**C**(O2), with a binding energy ranging from 41.25 to 27.48 kcal/mol. In their silver-equivalent **GC** hybrid structures, the order of stability follows **G**(N7)>**G**(N3)>**C**(O2)>**C**(O2 -nonplanar-) and the range of energy binding is from 33.44 kcal/mol to 26.59 kcal/mol. The binding between the obtained Au/Ag-**GC** neutral base pairs structures and their respective moieties, compared with the binding energy of the bare **GC** base pair shows a small change of around 2–3 kcal/mol in the binding energy upon addition of a neutral gold and silver atom. In the hybrid gold-**AT** neutral structures, the stability obeys the order **A**(N3)>**A**(N7)>**T**(O2-NH1)>**A**(N1)-**T**(NH3)>**T**(O4)>**T**(O2). In the silver equivalent hybrid **AT** structures, the order of stability follows **A**(N3)>**A**(N7)>**T**(O2-NH1)>**T**(O4)>**T**(O2 -nonplanar-)>**T**(O4 -nonplanar-). By comparing the results for the Au/Ag-**AT** hybrid system with its individual hybrid structures, we found no appreciable changes in the binding energies. In general, we can conclude that the addition of a neutral atom to the most stable formed base pairs does not produces perceptible changes in both structure and energetics in comparison with their equivalent WC pairs. Similar results have been found in previous studies [66].

The hybrid cationic gold-**GC**/**AT** shows a similar behavior to its related silvered systems. In both cases, the most stable structures are those ones where the metallic atom lies in between the two nucleobases. Additionally, there are more stable structures in the hybrid silver base pairs than in the gold structures. For the gold-**GC** structures, the order of stability follows the order **G**(O6)-**C**(N3)>**G**(N2)-**C**(N3)>**G**(N7). For this structures, the binding energy ranges between 151.77 and 117.09 kcal/mol. For the equivalent silver-**GC**, the stability obeys the order **G**(O6)-**C**(N3)>**G**(N7-O6)>**G**(N2)-**C**(O2)>**G**(N3)>**C**(O6)>**G**(O6) with the binding energy ranging from 111.05 kcal/mol to 63.34 kcal/mol. If we consider the isolated cationic Au/Ag-**GC** structures in comparison with their individual constituents, we found in the most stable configurations a reduction of around 10 kcal/mol in the base pair when both metals (gold and silver) bind the guanine nucleobase. In the hybrid gold-**AT** systems, the most stable structures are not planar, and they obey the stability **A**(N1)-**T**(O4)>**A**(N1-C2)-**T**(O2), with a binding energy from 141.22 kcal/mol to 106.29 kcal/mol. In the case of the silver-**AT**, the stability follows the order **A**(N1)-**T**(O4)>**A**(N1)-**T**(O2)>**A**(N6)-**T**(O4)>**A**(N3)>**A**(N7) where the binding energy ranges from 96.4 kcal/mol to 71.56 kcal/mol. The pairing effect reduces the binding energy in around 7 kcal/mol when the gold atom binds the adenine.

The anionic structures for both gold and silver atoms interacting with the **GC/AT** base pairs behave as the isolated nucleobases, that is, the metal atom binds the base pair only through the hydrogen atoms. The most stable structures in the hybrid anionic gold-**GC** systems follow the order of stability **C**(NH1)>**C**(NH5-CH6)>**G**(NH9)>**G**(NH2) with the binding energy ranging from 55.44 kcal/mol to 40.01 kcal/mol. In the respective silver structures, the order of stability obeys **C**(NH1)>**C**(CH6) and the binding from 53.39 kcal/mol to 48.75 kcal/mol. The change in the binding energy during the formation of the most stable metallic (gold and silver) anionic **GC** pair (in comparison with the independent parts) is an increase of around 5 kcal/mol in the cytosine and its respective loss in the guanine. In the anionic gold-**AT** structures, the order of stability is as follows: **T**(NH1)>**A**(NH9)>**A**(NH6) with an energy from 42.58 kcal/mol to 26.96 kcal/mol, and the respective order for the silver structures **T**(NH1)>**T**(CH6) with a binding energy from 40.07 kcal/mol to 28.82 kcal/mol. In this case, the variation in the energy binding due to pairing effect is <2 kcal/mol for both metals (gold and silver).

There is a big lack of scientific literature on the theoretical studies about the interaction of cationic and anionic silver and gold atoms with WC base pairs. The available results only consider the lowest energetic tautomers [69, 99]. Therefore, our results in energetics and stabilization can not be directly compared. In general, the interaction of metallic gold and silver atoms with WC base pairs follows the order in the binding energy: cationic>anionic>neutral. The obtained geometries and order of interaction presented in this review can be of great importance to predict the site of binding of more complex gold or silver metallic nanoclusters depending of the charge of the system.

#### 3.1.5 Deoxyguanosine monophosphate

The inclusion of the protonated sugar-phosphate backbone in the metal-guanine system brings new possible configurations. In the case of the neutral system, among the most stable structures are a set that is similar to the structures in the isolated metal-guanine case. In the most stable neutral structure for both metals, gold and silver bind the N7 site with the OH functional group. The stability in the gold case for the planar structures obeys the order N3>N7>O6 and for the silver N7>N3>O6 (we changed the order with respect to the isolated neutral Ag-guanine). The change in the energy binding in the comparable cases is a maximum of around 1 kcal/mol less when the sugar-phosphate backbone is considered.

In the cationic structures, the planarity is less favored due to a strong interaction of the metal with the oxygen atoms present in the phosphoric groups in the sugar backbone. In the cationic gold-**dGMP**, the most stable structure is nonplanar, and the order of stability follows N7>N3>O6. For the silver-**dGMP**, the most stable structure is planar and coincides with the isolated cationic silver-guanine N7-O6; the rest are nonplanar structures. For gold, the inclusion of the backbone in the cationic systems increases the binding energy by around 9–10 kcal/mol, whereas for silver is around 5–6 kcal/mol. In the anionic case, in both metals, a stronger interaction occurs with the OH functional groups present in the backbone. A comparison with the equivalent isolated cases shows that the binding energy for gold increases around 7–8 kcal/mol when the sugar backbone is included; meanwhile, in silver, it remains almost constant.

## 3.2 Ground state electronic properties

#### 3.2.1 Ionization potential and electron affinity

In the study of electronic properties of charged systems, it is important to start by calculating the ionization potential (IP) and electron affinity (EA) of the different species. The EA lets us estimate the anionic system, in which the two subsystems will most likely accept the extra charge. Using the IP, on the other hand, it is possible to estimate a possible charge redistribution during the cation's formation. Previous computational studies [67, 68, 98] have shown the importance to predict the charge transfer process during the formation of the hybrid metallic-DNA species.

In Table 2, we present the results in the calculation of the adiabatic, vertical IP, and EA as total energy differences and also from the energy of the highest occupied orbital for the DNA/RNA nucleobases, DNA base pairs, and the gold and silver atoms. The quantities were calculated using the definitions EA=E($N$)-E($N$+1) for the electron affinity and IP=E($N$-1)-E($N$) for the ionization potential. Here, E($N$) is the total ground-state energy of the neutral system and E($N$+1) and E($N$-1) the total energy in the same anionic (after adding one electron) or cationic (after removing one electron) system.

From the simulation, we can see that the EA of the metals is higher than all the nucleobases and base pairs. The results indicates then that the metal atom (Au/Ag) has a tendency to capture the electron during the formation of the anionic base-metal and pairs-metal structures (see Section 3.2.2).

The experimental values as well as the other theoretical calculations are all very close, with EA of metals higher than all the nucleobases, except for the EA of guanine, which can be smaller than Ag.

The subsystem with the lowest IP would most likely donate an electron to the bonded system. From the simulation, gold and silver have opposite characteristics. While the IP of Au is higher than all the nucleobases, the IP of Ag is smaller than all of them with the exception of guanine. From the simulated IP, we expect to see that in the cation's formation, the nucleobase loses one electron when bonded to Au. However, in the case of the Ag, we expect the opposite that the metal would lose the electron, again with the exception of guanine. In the experiment, the same trend is observed (the IP of Au higher than the IP of the nucleobases and the IP of Ag smaller than the IP of the nucleobases); however, there is no exception from guanine.

#### 3.2.2 Bader charge

The study of the redistribution of the electronic charge on hybrid charged metallic structures is of great value to understand the charge transfer, formation, and nature of the bonded structures. Even though other methodologies have been used: Mulliken analysis [52, 68] or natural bond orbital (NBO) analysis [53, 69, 110], in this work, we used the Bader charge population analysis method [81], previously used in the description of the charge transfer in the interaction of silver neutral clusters with nucleobases [64]. In our particular methodology, our results are free of the basis-set dependency issues of other methods [110].

**Table 2:** Adiabatic ionization potential ($IP_a$), vertical ionization potential ($IP_v$), adiabatic electron affinity ($EA_a$), and vertical electron affinity ($EA_v$) for the DNA/RNA nucleobases, WC base pairs, guanosine monophosphate and transition metals (gold and silver).

| Str. | $IP_a$ | $IP_v$ | $IP_{exp}$ | $EA_a$ | $EA_v$ | $EA_{exp}$ |
|---|---|---|---|---|---|---|
| G | 7.58 | 7.86 | 7.77–7.85[a], 8.0–8.3[b] | -0.027 | 0.069 | -0.44[c], -2.07 to -0.08[d] |
| | 7.81[e], 7.51[f] | 8.05[e], 7.80[f] | | -0.049[e], -0.025[f] | | |
| A | 8.00 | 8.16 | 7.80–8.55[a], 8.3–8.5[b] | -0.279 | 0.317 | -0.72[c], -0.56 to -0.45[d] |
| T | 8.61 | 8.78 | 8.80–8.87[a], 9.0–9.2[b] | -0.043 | 0.058 | 0.02–0.068[c], -0.53 to -0.29[d] |
| C | 8.63 | 8.59 | 8.45–8.68[a], 8.8–9.0[b] | -0.107 | 0.187 | -0.10[c], -0.55 to -0.32[d] |
| U | 9.11 | 9.22 | 9.20–9.32[a], 9.4–9.6[b] | -0.079 | 0.018 | 0.03[c], -0.30 to -0.22[d] |
| GC | 6.82 | 7.11 | 7.05, 7.50[g] | 0.184 | -0.184 | -0.48[h] |
| AT | 7.43 | 7.53 | 7.85, 8.27[g] | 0.287 | -0.179 | -0.18[h] |
| dGMP | 7.32 | 7.75 | 7.96[i] | 0.31 | 0.24 | 0.24, 0.14[j] |
| Au | 9.54 | – | 9.23[k] | 2.25 | – | 2.31[k] |
| Ag | 7.91 | – | 7.58[k] | 1.22 | – | 1.30[k] |

[a]Experimental adiabatic values in TABLE IV in Ref. [104].
[b]Experimental vertical values in TABLE I in Ref. [104].
[c]Theoretical adiabatic values in TABLE III in Ref. [105] from CASPT2(IPEA)/ANO-L 4321/321//CCSD/aug-cc-pVDZ calculations.
[d]Experimental vertical values in TABLE II in Ref. [105] except for guanine where the values are from the B3LYP range.
[e]With LDA.
[f]With RPBE.
[g]ZPE-corrected M06-2X/6-31++G(d,p). TABLE II in Ref. [106].
[h]ZPE-corrected M06-2X/6-31++G(d,p). TABLE I in Ref. [106].
[i]Vertical IP at MP2/6-311G(d,p)//P3/6-311G(d,p) level. TABLE III in Ref. [107].
[j]Adiabatic and vertical values at B3LYP/DZP++ level. TABLE I in Ref. [108].
[k]From Ref. [109].

As expected from the high values of the EA of the metals in the case of the anionic metal-nucleobase structures, there is a large electronic density accumulated in the metal atom. The Bader charge varies only a little among all obtained single and paired structures (from -0.8 $|e|$ to -0.9 $|e|$, see Table 3). The extra charge present in the total system is practically localized in the metal atom.

In the cationic case, the missing electronic charge density is removed mainly from the metal atom. However, important variations reflect the difference in the IP of Au and Ag. The positive Bader charge of Au is smaller than the positive Bader charge of Ag (it varies from 0.4 $|e|$ to 0.6 $|e|$ and from 0.7 $|e|$ to 0.8 $|e|$, respectively). No exceptional value is found in guanine, despite a different simulated IP. In the cationic case, even if most of the electronic charge is removed from the metal, there is an important contribution from the nucleobase. A trend is observed, wherein the metal retains a higher charge (in anion) or loses less charge (in cation) when the metal-nucleobase bond involves a single oxygen atom; this explains the Bader's variability for a given nucleobase. The structures with the oxygen-metal bond are formed with the thymine and uracil nucleobases.

**Table 3:** Bader analysis on the metal atom (in *e* units): gold and silver (and HOMO-LUMO gap) neutral[a] for the most stable DNA/RNA hybrid metal structures obtained at PBE level.

| System | Gold (79) | | | Silver (47) | | |
|---|---|---|---|---|---|---|
| | Neutral | Cation | Anion | Neutral | Cation | Anion |
| G | -0.2 (2.05/2.65) | 0.5 (1.23) | -0.8 (2.23) | -0.1 (3.28/1.23) | 0.7 (2.30) | -0.9 (1.35) |
| A | -0.2 (2.43/1.86) | 0.4 (1.28) | -0.9 (1.80) | -0.1 (3.20/1.28) | 0.7 (1.49) | -0.9 (0.94) |
| T | -0.1 (1.86/1.85) | 0.6 (1.31) | -0.8 (1.77) | 0.0 (2.88/1.60) | 0.8 (1.70) | -0.8 (0.88) |
| C | -0.2 (2.43/1.22) | 0.5 (1.94) | -0.9 (1.71) | -0.1 (3.27/0.67) | 0.7 (2.81) | -0.9 (0.80) |
| U | -0.1 (1.85/1.75) | 0.6 (1.57) | -0.8 (1.77) | 0.0 (2.91/1.62) | 0.8 (1.96) | -0.8 (0.83) |
| GC | -0.2 (2.01/1.49) | 0.4 (2.47) | -0.9 (1.64) | -0.1 (2.63/0.33) | 0.6 (1.87) | -0.8 (0.71) |
| AT | -0.2 (2.40/1.63) | 0.4 (3.06) | -0.9 (1.11) | -0.1 (3.11/1.14) | 0.6 (3.12) | -0.8 (0.29) |
| dGMP | -0.1 (2.26/2.25) | 0.3 (2.26) | -0.7 (2.11) | 0.0 (2.82/1.71) | 0.7 (2.24) | -0.7 (1.12) |

[a]For neutral systems with HOMO-SOMO/SOMO-LUMO gaps.

In the neutral system, almost no charge is present in the formed metal-nucleobase, in accordance with the very low binding energy. A small negative Bader charge is present in the metal atom in the neutral metal-nucleobases structures.

The system that includes **AT** and **GC** base pairs have a similar metal atom Bader charge, again with variations of the order of 0.1 |*e*|. In the anionic case, the metal is bound to one of the pairs, and the Bader charge is -0.8 |*e*| and 0.9 |*e*| for Ag and Au, respectively. In the cation, the metal has a Bader charge of 0.6 |*e*| and 0.4 |*e*| for Ag and Au.

The addition of backbone to the system guanine-metal changes the Bader charge by 0.2 |*e*| in the most extreme case. The metal accumulates most of the charge in the anionic case, and the Bader charge is -0.7 |*e*| for both Au and Ag, but this is smaller in absolute value than the Ag-**G** alone by 0.2 |*e*|. Indeed, the bonding configuration is completely different with the backbone atoms participating in it. In the case of the cation, Ag has the same Bader charge, but Au has a Bader charge of 0.3 |*e*| that is 0.2 |*e*| lower than Ag-**G** alone. Again, the configurations are different with the participation of the backbone atoms in the bonding.

In general, our results agree with similar previous studied systems [57, 68, 70]. The nature of the metallic atom more than the nucleobase drives the charge transfer (donated or accepted) depending of the overall charge in the hybrid geometry. These effects can be of importance in the design of electronic and optical tuned devices.

### 3.2.3 Electronic gap

Underestimation of the fundamental electronic gap with local and gradient-corrected density functional theory methods is a well-known issue in computational material sciences [111, 112]. In our case, the additional charge transfer process due to the metallic hybridization can present an extra challenge that can be corrected with long-range corrected functionals [113, 114]. Despite these considerations, in this section, we summarized the electronic HOMO-LUMO gaps with the hope that the consigned values will help as a first step to predict general tendencies in the electronic excitations of the studied systems (See Table 4).

The electronic HOMO-LUMO gap varies importantly among the structures. We, therefore, can expect a high dependence on the system's optical properties to discern the metal-attaching site inside a DNA strand. Not all cases could be (in principle) distinguished by the use of the electronic gap, but it will be a powerful tool as a guide to rule out configurations.

For a given metal, the simplest case to consider is the anion. The HOMO-LUMO gap of the formed Au-nucleobases is 1.7-1.8 eV in the case of the single bases (**A**, **T**, **C**, **U**). The Au-guanine gap is strikingly different (2.23 eV), it is reduced by 0.1 eV when the backbone participates in **dGMP** system. In the case of the metal bonded to the pair **AT**, the gap is reduced considerably (by 0.6 eV) with respect to the corresponding Au-nucleobases.

**Table 4:** Electronic HOMO-LUMO[a] (in eV) gaps of all DNA/RNA hybrid metal structures studied in this work (from the most to the least stable) at the PBE level.

| System | Au | | | Ag | | |
|---|---|---|---|---|---|---|
| | Neutral | Cation | Anion | Neutral | Cation | Anion |
| G | 2.05, 2.28, 1.77, 1.56, 0.35, 0.25 | 1.23, 0.56, 0.43 | 2.23, 1.61, 1.92 | 3.28, 2.43, 2.45, 0.67 | 2.30, 0.80, 0.93, 0.82, 0.71 | 1.35, 0.74 |
| A | 2.43, 2.43, 2.41, | 1.28, 1.26, 0.81 | 1.80, 1.28, 1.48 | 3.20, 3.19, 2.95 | 1.49, 1.51, 1.12 | 0.94, 0.47 |
| T | 1.86, 1.80, 1.78, 0.58 | 1.31, 0.81, 0.97 | 1.77, 0.96 | 2.88, 3.12, 3.00 | 1.70, 1.04, 1.44, 1.28 | 0.88 |
| C | 2.43, 1.94, 1.72, | 1.94, 1.95, 1.57 | 1.71, 1.69, 1.41 | 3.27, 3.02, 3.09 | 2.81, 1.91 | 0.80, 0.81, 0.48 |
| U | 1.85, 1.76, 1.73, 1.76 | 1.57, 1.37, 0.91, 1.11 | 1.77, 0.93 | 2.91, 3.10, 3.15, 3.08 | 1.96, 1.79, 1.31, 1.40, 1.33 | 0.83 |
| GC | 2.01, 2.32, 0.94, 1.67, 3.43, 1.33 | 2.47, 2.40, 0.75 | 1.64, 1.35, 1.09 0.37 | 2.63, 2.39, 1.59, 1.72 | 1.87, 1.72, 3.45, 0.66, 0.03, 0.15 | 0.71, 0.31 |
| AT | 2.40, 2.45, 1.58, 2.39, 1.60, 1.55 | 3.06, 2.01 | 1.11, 1.09, 0.25 | 3.11, 2.95, 2.23, 2.33, 2.44, 1.93 | 3.12, 3.53, 3.40, 0.44, 0.12 | 0.29, 0.08 |
| dGMP | 2.26, 1.75, 1.58, 2.28, 1.79, 1.70 | 2.26, 1.21, 2.12, 1.51, 0.69, 0.61 | 2.11, 2.04, 1.69, 1.71, 1.61, 1.22 | 2.82, 3.29, 2.22, 2.42, 1.86, | 2.24, 2.20, 2.62, 2.32, 1.44 | 1.12, 1.20, 0.87, 0.76,0.59, 0.42 |

[a]For neutral systems the HOMO-SOMO gap.

The same trend is valid for silver. The Ag-**G** gap is the highest (1.35 eV), reduced by 0.2 eV with regard to participation of backbone. All the other Ag-nucleobases (**A**, **T**, **C**, **U**) have a similar gap, but they are more spread out than the Au case (from 0.80 to 0.94 eV). In the instance where Ag is bonded to the pair **AT**, the gap is again reduced considerably (by more than 0.6 eV) with respect to the corresponding Ag-nucleobases. In the anion case, the systems's gaps are smaller when the metal is Ag than it is with Au.

In the case of the cation, the opposite effect is observed. the Ag-bases gap is higher than the Au-bases gap. The electronic gap in the Au-pair system varies greatly for the single bases (from 1.23 eV to 1.94 eV). The case of metal bonded to the pair **GC** and **AT** changes the gap. It is increased considerably (by more than 0.5 eV and 0.7 eV for **GC** and **AT**, respectively). Including the backbone in the bonding to Au also drastically changes the electronic gap. Ag-**G** changes from 1.23 eV to 2.26 eV in **dGMP**. The case of the cation and Ag is similar when considering the gap variability within the single bases (from 1.49 to 2.81 eV). The gap in the metal-nucleobase pairs is completely different than the single ones. Contrary to Au, the inclusion of backbone does not significantly change the gap between M-**G** and **dGMP**.

As we can see, there are several trends in the electronic gap for the studied systems with a strong dependency of the nucleobase, charge, or metal. Therefore, a more detailed study of the optical properties, for example, using time–dependent density functional theory is desired. These aspects will be addressed by the authors in a future work.

### 3.2.4 Frontier Kohn-Sham orbitals

The form of the frontier HOMO-LUMO orbitals has the expected shape, following the analysis of the Bader charge. The shape of the orbitals for the different bases and metals are similar, and we include the adenine case in Figure 6 as a reference.

In the case of the anions, the extra electronic charge is highly localized in the metal. The HOMO of the anionic case is also localized in the metal and corresponds as expected to the metals' closed 5*s* or 6*s* electronic shell. The anionic system's LUMO is in all cases the LUMO of the nonbonded nucleobase.

In the case of the cationic system, the electronic charge is donated by both subsystem metal and the nucleobase. Correspondingly, the form of the LUMO in the cationic case is an electron delocalized between the metal and the nearest atom(s) in the nucleobase forming the bond.

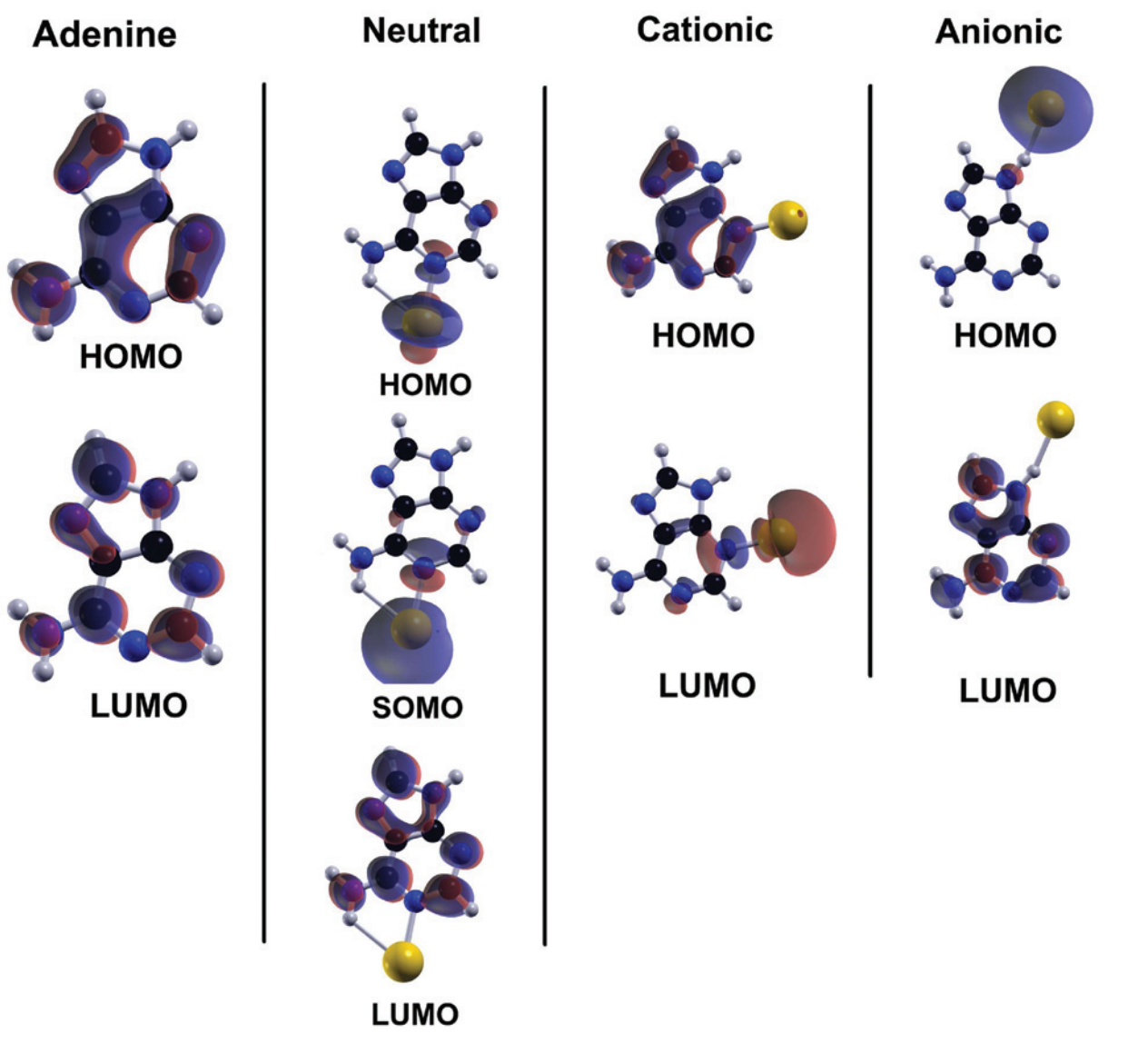

**Figure 6:** Representative HOMO-LUMO Kohn-Sham orbitals for the isolated adenine (**A**) nucleobase (on the *left*) and its lowest energetic gold hybrid structures in three charge states: neutral, cationic, and anionic.

Because there are different types of atoms in the bond between the structures, this explains the variability of the electronic gap. The HOMO of the cationic system is the HOMO of the nonbonded nucleobase.

When forming the pairs (see Figure 7), the HOMO of the anion remains localized in the metal and LUMO in one of the bases. In the pair **AT**, while the metal binds to **T**, the LUMO is localized in **A**. The opposite appears in the **GC** pair. The metal binds to **C**, but the LUMO is localized in **C**. In the cation, where the metal is binding simultaneously both bases in the pair, the HOMO and LUMO are localized in the bases. In the **GC** pair, HOMO is in **G**, while LUMO is in **C**; and in the **AT** pair, HOMO is in **A**, while LUMO in **T**.

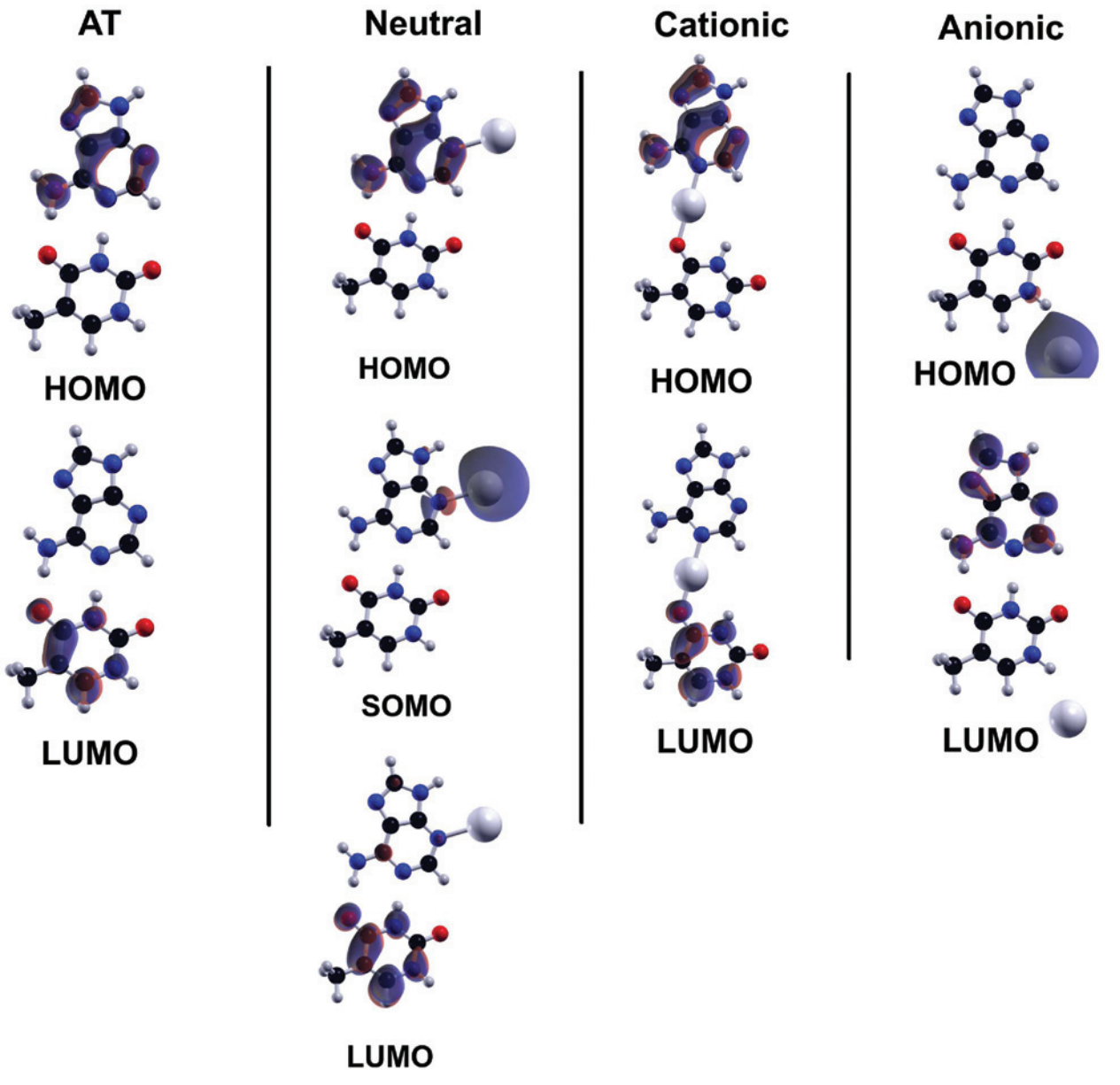

**Figure 7:** Representative HOMO-LUMO Kohn-Sham orbitals for the isolated adenine-thymine (**AT**) base pair (on the *left*) and its lowest energetic silver hybrid structures in three charge states: neutral, cationic, and anionic.

The anion system that includes the backbone has the same HOMO and LUMO as the metal-base alone (see Figure 8). The cation case, however, is quite different, as would be expected from a new configuration that includes a bond between the metal and atoms in the backbone. The HOMO is delocalized between the metal and the base (it was only on the base in the absence of the backbone), and the LUMO is also an orbital delocalized between the metal and the base. Depending on the case, the HOMO-LUMO transition is expected to have a low or high oscillatory strength in optical absorption. A complete analysis of the other orbitals, along with their relation to bonding and optical properties, will be reported in a future study.

## 4 Conclusions

The computational study of the interaction of gold and silver atoms with DNA/RNA has been an active subject of research in the last 20 years. In this work, we have presented the historical path of the first attempts to describe the mechanisms of formation and the electronic properties of the hybrid structures complemented with our own calculations using a density functional theory methodology. This combination provides an exhaustive study of the interaction, stability, and electronic properties of the DNA/RNA nucleobases interacting with the noble metal atoms (gold and silver) in three charge states: cationic, neutral, and anionic using a DFT real-space methodology implementation. We have taken into account the paring effect by studying the WC base pairs and by including the sugar backbone in the guanine nucleobase.

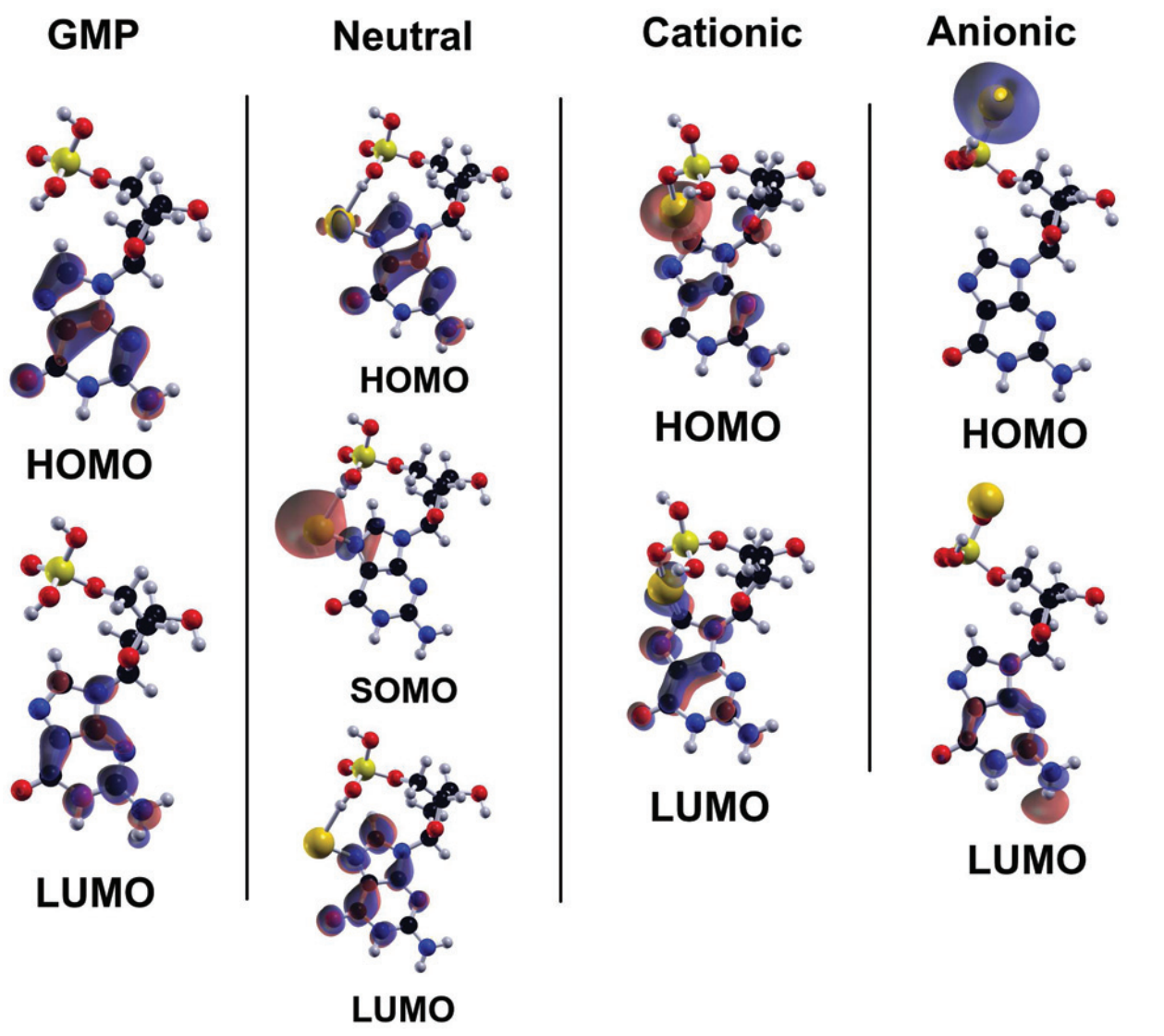

**Figure 8:** Representative HOMO-LUMO Kohn-Sham orbitals for the isolated guanine monophosphate (**dGMP**) structure (on the *left*) and its lowest energetic gold hybrid structures in three charge states: neutral, cationic, and anionic.

We found that the hybrid metal structures topology is dominated by an electronic redistribution of charges in the molecule in accordance with previous studies. In general, nitrogen, oxygen, carbon, and phosphorus behave as negative centers, while hydrogen behaves as a positive center. Comparing gold and silver hybrid geometries, we can conclude that the bond lengths in the hybrid DNA/RNA-silver structures are longer than their gold counterparts independently of the charge state. In the neutral structures, the metal binds the nucleobase through both negative and positive centers, where the *s* electronic orbital in the metal atom hybridizes with the *d* orbital by redistributing the charge to favor the dipolar interaction. Moreover, in the neutral case, the very small energy difference between the most stable geometries can induce a change in the relative ordering due to the exchange-correlation functional inaccuracy. In the charged structures, the strong electrostatic interaction produces the same low-energy hybrid geometries for both metallic atoms, however, the ordering of the less stable structures can change. The set of structures obtained through the interaction of the metallic atoms with WC pairs are diverse in relation with their moieties. This can be related to the presence of more centers for binding. In the neutral and anionic structures, the most stable geometries show the metallic atom binding the system in regions far from the zone where the hydrogen bonds are present. In both cases, the interaction with the metallic atom do not change substantially the geometry of the pair in the most stable systems. On the contrary, the most stable cationic geometries are those where the metallic atom binds the WC moieties in the zone of the hydrogen bonds, changing considerably the geometry of the base pair. The less stable cationic geometries tends to preserve the geometry of the WC pair. The inclusion of the sugar-phosphate backbone in the nucleobase opens the possibility for new binding sites, in particular, in the anionic structures. The most stable hybrid gold neutral and cationic structures tend to keep the binding sites in the nucleobase in combination with the binding sites in the phosphate.

For anions and cations, the bonding energy of the metal-base increases as the Bader charge of the atom base increases, and the bonding energy decreases when atoms with opposite charge centers are present. The Bader analysis showed that the gain in electronic charge is mainly localized on the noble metal atom for anions. For cations, the electronic charge is donated partially by the metal and partially by the nucleobase (so the donation charge is shared almost equally).

The frontier orbitals in the neutral structures show the induced hybridization by the binding. For the anionic case, the orbital analysis of frontier orbitals is also homogeneous. The HOMO orbitals are localized in the metal, and the LUMO are localized in the nucleobase, with no change induced by the inclusion of the backbone or pairing. In the cationic case, the HOMO is localized in the base, and also in the backbone when included, while the LUMO is delocalized between the metal and the base. When considering pairs, HOMO and LUMO are each localized in a different nucleobase. Finally, the electronic gap varies greatly among all of the considered structures and is particularly sensitive to the backbone participation in the bonding. Thus, it could be used as a fingerprint when searching Au/Ag-DNA hybrid atomic structures.

Despite the progress in both the computational and experimental areas, the study of the interaction of gold and silver with DNA/RNA is still a fruitful field of research. Interesting phenomena such us stabilization, geometry, fluorescence, conductivity among others still remains unexplained or are not well understood. In this work we compile, summarize, and include as much computational information as possible about the ground state properties of hybrid gold and silver atoms with DNA/RNA nucleobases as a solid starting point to future research in the area where more complex systems will be considered.

## 5 Associated content

Supporting Information supplies a comparison of Guanine-Au and Guanine-Ag electronic properties obtained with different exchange-correlation DFT functionals and structural information of the most stable structures (up to six) for the three charge states: neutral (0), cationic (+1), and anionic (-1) in both hybrid Au/AG-DNA/RNA systems.

**Acknowledgments:** We gratefully acknowledge financial support from the Academy of Finland via the Centre of Excellence in Computational Nanoscience (COMP) project numbers 251748 and 279240; and we are grateful to CSC, the Finnish IT Center for Science in Espoo, for computational resources.

**Supplemental Material:** The online version of this article (DOI: 10.1515/ntrev-2012-0047) offers supplementary material, available to authorized users.

## Bionotes

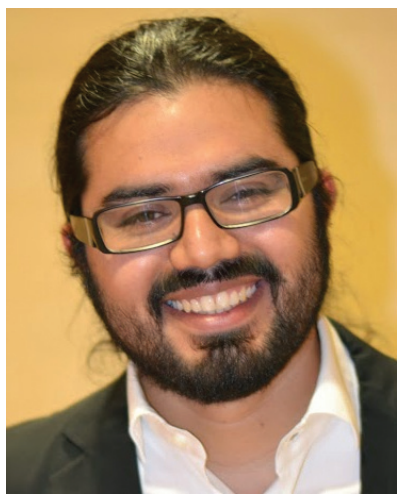

**Leonardo Andres Espinosa Leal**
COMP Centre of Excellence, Department of Applied Physics, Aalto University, P.O. Box 11100, 00076 Aalto, Finland

Leonardo Andres Espinosa Leal received his degree in physics from the National University of Colombia in 2005. He obtained his Master's degree in nanoscience in 2009 and his PhD (2013) in physics of nanostructures and advanced materials (PNAM), under the supervision of Prof. Angel Rubio at the University of the Basque Country (Spain). He is currently a postdoctoral researcher in the Computational Soft and Molecular Matter (CSMM) group led by Dr. Olga Lopez Acevedo in the Centre of Excellence on Computational Nanosciences, Aalto University, Finland. His research interests are the electronic properties in the ground and excited state of biological systems.

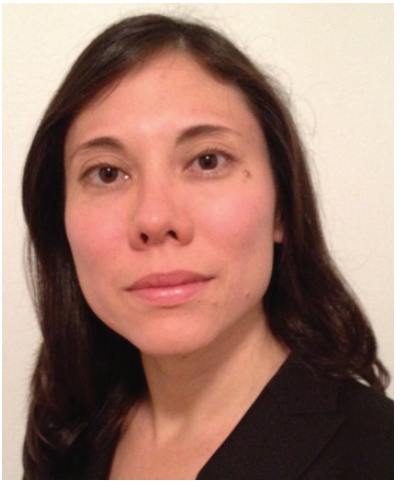

**Olga Lopez-Acevedo**
COMP Centre of Excellence, Department of Applied Physics, Aalto University, P.O. Box 11100, 00076 Aalto, Finland,
**olga.lopez.acevedo@aalto.fi**

Olga Lopez-Acevedo is an Academy Fellow and Group Leader in the Centre of Excellence on Computational Nanosciences, Aalto

University, Finland. Her research interest focuses on the quantum modeling of hybrid soft-nano systems and the development of computational methods for multi-scale simulations. She obtained her Bachelor's and Master's degrees in physics from the University of Strasbourg, France. In 2006, she obtained a PhD in theoretical physics working on quantum algorithmics from the University of Cergy-Pontoise, France. Her postdoctoral experience includes stays at the Université Libre de Bruxelles, Belgium, and University of Jyväskylä, Finland.